\documentclass[12pt,a4paper,english]{article}
\usepackage{titling}
\usepackage[affil-it]{authblk}
\usepackage[dvipsnames]{xcolor}
\usepackage{longtable}
\usepackage{lscape}
\usepackage{graphicx}
\usepackage{epsfig}
\usepackage{dsfont}
\usepackage{amsfonts}
\usepackage{amsmath}
\usepackage{amsthm}
\usepackage{amssymb}
\usepackage{bbold}
\usepackage[english]{babel}
\usepackage[utf8]{inputenc}
\usepackage{calc}
\usepackage{cancel}
\usepackage{float}
\usepackage{slashed}
\usepackage{mathrsfs}
\usepackage{pdfpages}
\usepackage{tabu}
\usepackage{simplewick}
\usepackage[hidelinks]{hyperref}
\usepackage{multirow}
\usepackage{hhline}
\usepackage{float}
\usepackage{bm}
\usepackage{braket}
\usepackage{comment}
\usepackage{cite}
\usepackage{soul}
\usepackage[left=0.8in,right=0.8in,top=1.0in,bottom=1.2in]{geometry}
\usepackage{duckuments}

\usepackage{datetime}
\usepackage{booktabs}
\date{}

\numberwithin{equation}{section}
\numberwithin{figure}{section}
\numberwithin{table}{section}

\makeatletter
\g@addto@macro\bfseries{\boldmath}
\makeatother

\allowdisplaybreaks

\newcommand{\published}[1]{%
\gdef\puB{#1}}
\newcommand{\puB}{}

\begin{document}

\title{\textbf{Implications of $K\to\pi\nu\bar\nu$ for new physics in $B$ decays}}
\author[1]{Lukas Allwicher \thanks{lukas.allwicher@desy.de}}
\author[2]{Marzia Bordone\thanks{marzia.bordone@uni-mainz.de}}

\affil[1]{Deutsches Elektronen-Synchrotron DESY, Notkestr. 85, 22607 Hamburg, Germany}
\affil[2]{\textit{PRISMA}$^{++}$ Cluster of Excellence \& Mainz Institute for Theoretical Physics
Johannes Gutenberg University, Staudingerweg 9, D-55128 Mainz, Germany}

\published{\flushright 
MITP-26-035\\
DESY-26-102
\vskip2cm }

\maketitle
\begin{abstract}
The new measurement of the $K^+\to\pi^+\nu\bar\nu$ branching ratio by the NA62 collaboration represents an important milestone in precision flavour physics.
Under different flavour assumptions on beyond the standard model physics, this decay mode can be related to many other observables with underlying flavour-changing  currents, such as $B\to K\nu\bar\nu$ or $B_s\to\mu\mu$, effectively providing an additional handle in our exploration of flavour-violating new physics effects.
We study the impact of this new result, under the hypotheses that deviations from the standard model arise from modified $Z$ couplings, or from effects in semileptonic interactions.
Finally, we review the implications of our scenarios for the yet to be measured partner mode $K_L\to\pi^0\nu\bar\nu$.
\end{abstract}
\newpage

\section{Introduction}
Flavour-changing neutral-current (FCNC) processes remain among the most sensitive probes of physics beyond the Standard Model (SM).
Their strong suppression within the SM, due to the GIM mechanism, makes them very rare and, therefore, particularly sensitive to both heavy and light new degrees of freedom. Among these processes, the $K^+\to\pi^+\nu\bar\nu$ decay mode plays a special role: its exceptionally small branching ratio, of $\mathcal{O}(10^{-10})$, with theoretical uncertainties well under control, provides a clean test of flavour violation involving the first two quark generations. When complemented by other theoretically clean FCNC modes, such as $B_s\to\mu^+\mu^-$ and $B\to K\nu\bar\nu$, it provides a precise probe of the flavour structure of possible New Physics (NP) effects across different quark generations.

On the experimental side, the branching fraction of $B_s\to\mu^+\mu^-$ is measured by ATLAS, CMS and LHCb, with a combined experimental uncertainty of roughly $8\%$ \cite{LHCb:2021awg,LHCb:2021vsc,CMS:2022mgd,Greljo:2022jac}. The first evidence of the $B^+\to K^+\nu\bar\nu$ decay has been recently reported by the Belle II experiment \cite{Belle-II:2023esi}, with a branching fraction mildly above the SM. Finally, the NA62 experiment has recently released a new measurement of the $K^+\to\pi^+\nu\bar\nu$ decay, with the available datasets up to 2024. This is one of the most precise measurements of a rare decay as of yet, and it is found to be in agreement with the SM \cite{NA62:2026rwr}. The recent progress in $K^+\to\pi^+\nu\bar\nu$ is particularly timely. 
Reaching a relative uncertainty below $20\%$ in this mode, together with increasingly precise measurements in $B$ physics, allows us to quantitatively test flavour hypotheses that connect flavour violation in the light-quark families to the third generation. The aim of this work is to quantify the impact of $K\to\pi\nu\bar\nu$ measurements, and to assess their complementarity with other flavour and electroweak observables in motivated NP scenarios.

The interplay of these modes has already been studied in the literature \cite{Abada:2026dlb,Das:2026hpy,Crivellin:2025qsq,Allwicher:2024ncl,Buras:2024ewl,Marzocca:2024hua,Marzocca:2021miv,Blanke:2026rdi}. In this work, we reassess these correlations using the latest
$K^+\to\pi^+\nu\bar\nu$ measurement and compare their impact under different
assumptions on the flavour structure of NP.
We consider two complementary scenarios with heavy NP, together with motivated hypotheses on their flavour structure. First, we discuss the possibility that NP affects only quark couplings. In this case, NP effects can be encoded in modified $Z$ couplings, allowing us to compare the resulting low-energy bounds with those from electroweak precision observables. This setup can be viewed as a proxy for scenarios in which NP couples directly to operators involving two quark fields and a Higgs current. For this scenario, we consider two flavour assumptions: Minimal Flavour Violation \cite{DAmbrosio:2002vsn} and Partial Compositeness \cite{Gherghetta:2000qt,Agashe:2004cp,Contino:2006nn,Davidson:2007si,Keren-Zur:2012buf}. These two flavour hypotheses allow us to test the complementarity of these measurements: while in the former we see that, due to the alignment of NP to the SM flavour structure, $B_s\to\mu^+\mu^-$ is the most constraining FCNC process so far, in the latter we find that $K^+\to\pi^+\nu\bar\nu$ is essential to disentangle possible solutions for the NP couplings. \\
Second, we consider the complementary possibility that NP appears in dimension-six four-fermion semileptonic operators, thereby extending the analysis beyond FCNC observables to charged-current transitions.
 We work with a $U(2)^5$ flavour symmetry, which effectively imposes a third-generation dominance in the NP couplings \cite{Barbieri:2011ci}. This choice is motivated by the current pattern of data, where several measurements involving third-generation fermions show tensions with the SM, and are well accommodated in this framework \cite{Allwicher:2024ncl,Crosas:2022quq}. We build on the previous analysis of  Ref.~\cite{Allwicher:2024ncl}, updating it to the most recent experimental inputs, when needed. Here we find an interesting relation between the $K^+\to\pi^+\nu\bar\nu$ and the $K_L\to\pi^0\nu\bar\nu$, which predicts their relative size as a function of the fit parameters, and provides a further complementary test of this scenario.

The paper is organised as follows. We start in Section \ref{sec:Z} by analysing the case in which the source of NP in FCNCs is given by modified $Z$ couplings. In Section \ref{sec:U2} we present the current status of a global analysis with NP in semileptonic interactions, under the assumption of third-generation dominance. After that, we review the implications for phenomenology and the still unmeasured $K_L\to\pi^0\nu\bar\nu$ branching ratio (Section \ref{sec:KL}) before concluding.

\section{FCNCs from effective $Z$ couplings}\label{sec:Z}

In this section we investigate the possibility of having new physics in flavour-violating $Z$-couplings to down-type quarks, following the discussion in~\cite{Guadagnoli:2013mru}. This setup makes the correlation between rare flavour-changing modes particularly transparent: once a flavour-changing $Z\bar{d}_i d_j$ vertex is present, it induces tree-level contributions to both charged-lepton and dineutrino FCNC processes. The relative impact in $B_s\to \mu^+\mu^-$, $B\to K\nu\bar\nu$, and $K\to\pi\nu\bar\nu$ is then controlled by the flavour structure of the non-standard $Z$ couplings, while the lepton couplings are fixed to their SM values, and are, therefore, universal in the three families.

At the electroweak scale, we can work with the simplified Lagrangian
\begin{equation}
\mathcal{L} = \frac{g}{c_W} Z_\mu \bar{d}_i \gamma^\mu \left[(g_L^{ij}+\delta g_L^{ij})P_L+(g_R^{ij}+\delta g_R^{ij})P_R\right] d_j\,,
\end{equation}
where $g$ is the weak $SU(2)_L$ coupling, and $c_W(s_W)$ is the cosine(sine) of the weak angle. At tree-level in the SM, we have that
\begin{align}
g_{L}^{ii}=\, -\frac{1}{2}+\frac{1}{3}s_W^2\,, \quad g_{R}^{ii}=\, \frac{1}{3}s_W^2\,, \quad g_{L(R)}^{ij}=0\quad \mathrm{for}\quad i\neq j\,.
\end{align}
The non-standard contributions scale differently for our two flavour assumptions, namely
\begin{align}
    (\delta g_L^{ij})^{\rm MFV} &= \frac{V_{ti}^*V_{tj}}{|V_{tb}|^2} \delta g_L \,, \qquad &(\delta g_R^{ij})^{\rm MFV} &= \frac{m_{d_i}m_{d_j}}{m_b^2} \frac{V_{ti}^*V_{tj}}{|V_{tb}|^2} \delta g_R \,, \\
    (\delta g_L^{ij})^{\rm PC} &= \frac{|V_{ti}| |V_{tj}|}{|V_{tb}|^2} \delta g_L \,, \qquad &(\delta g_R^{ij})^{\rm PC} &= \frac{m_{d_i}m_{d_j}}{m_b^2} \frac{|V_{tb}|^2}{|V_{ti}| |V_{tj}|} \delta g_R
\end{align}
for Minimal Flavour Violation (MFV) and Partial Compositeness (PC) respectively. The normalisation is chosen such that $ (\delta g_{L(R)}^{33}) = \delta g_{L(R)}$, both in the MFV and PC case.\\
The two flavour hypotheses lead to rather different phenomenological patterns. In MFV, flavour-changing couplings are aligned with the CKM structure of the SM. Moreover, right-handed contributions are additionally suppressed by light-quark masses, making them particularly small in transitions involving the first two generations. In the PC-inspired scaling, instead, the left-handed couplings follow a CKM-like hierarchy, while the right-handed ones exhibit a different parametric behaviour: the inverse CKM factors partly compensate the light-quark-mass suppression. As a result, Kaon observables can become much more sensitive to $\delta g_R$ than in the MFV case, as we shall show in the following.

\begin{figure}
    \centering
    \includegraphics[width=0.49\linewidth]{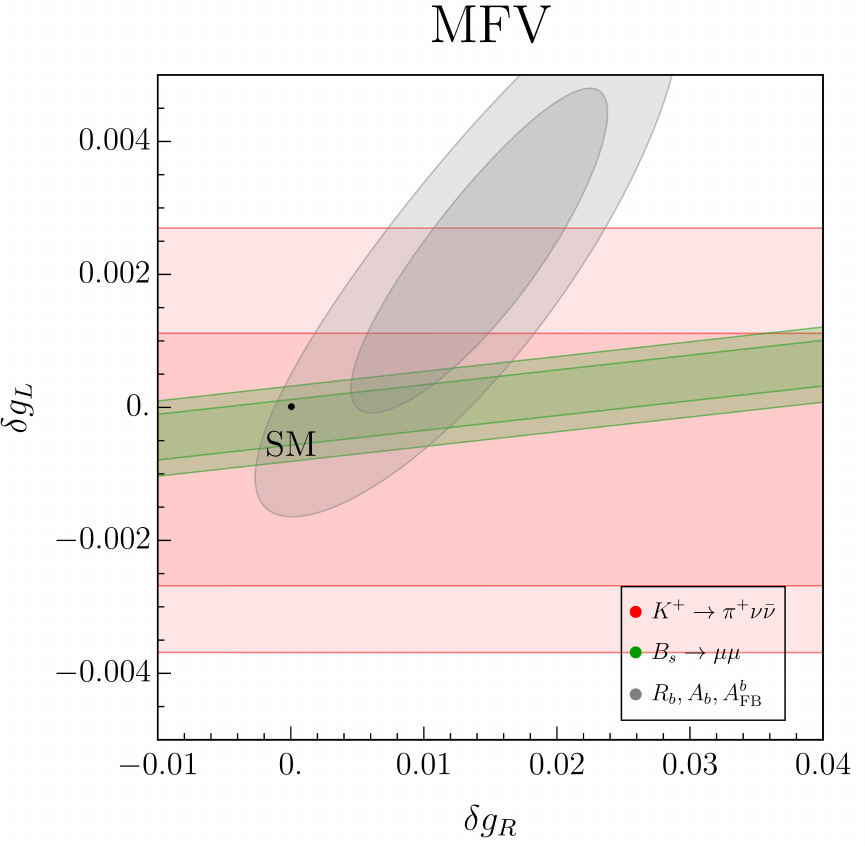}
    \includegraphics[width=0.49\linewidth]{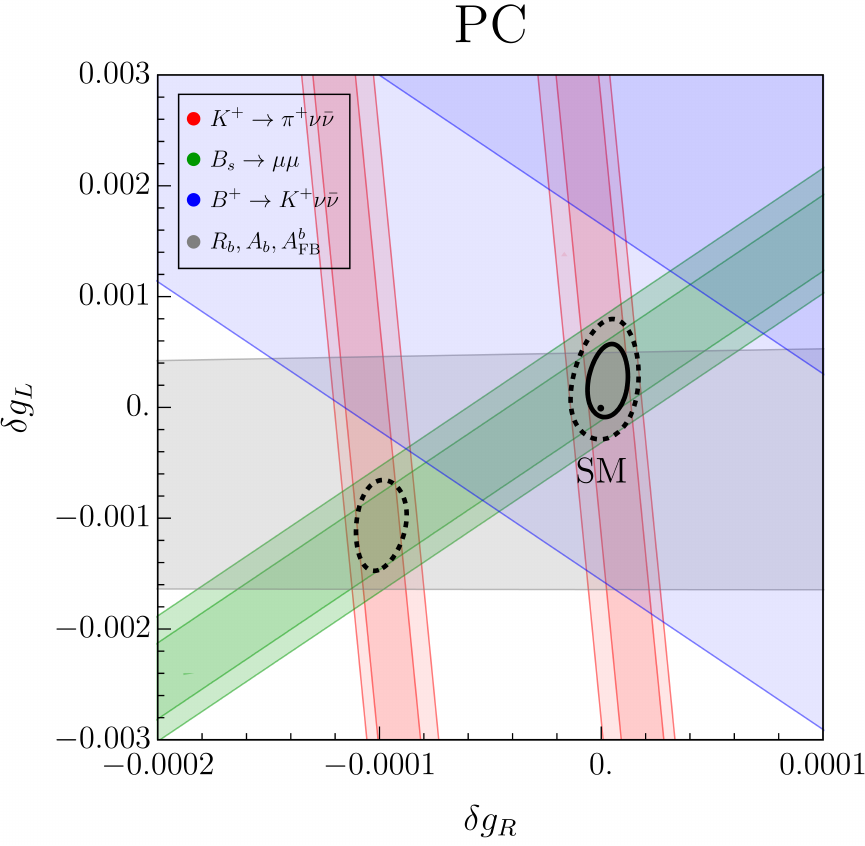}
    \caption{Constraints on effective $Z$ couplings under the assumption of Minimal Flavour Violation (left) and Partial Compositeness (right). In grey the $1$- and $2\sigma$ allowed region from the combined $Z\to b\bar b$ observables, in green from $B_s\to\mu\mu$, and in red from $K\to\pi\nu\bar\nu$. In the right plot, the blue regions show the preferred regions from $B\to K\nu\bar\nu$, while the black solid (dashed) lines are the $1$- and $2\sigma$ contours of the full combined fit.}
    \label{fig:Zcouplings}
\end{figure}

The main flavour-conserving constraints on $\delta g_{L(R)}$ come from electroweak precision tests, in particular $Z\to b\bar b$ decays measured at LEP, and usually condensed in the observables $R_b$, $A_b$, and $A_{\rm FB}^b$.
We show these constraints as grey regions in Figure \ref{fig:Zcouplings}.
On the flavour-violating side, the allowed parameter space varies with the flavour assumption.
We observe that:
\begin{itemize}
    \item[\textbf{MFV}] In this case, the right-handed coupling is suppressed by light quark masses, $m_s/m_b$ for the $B_s$ case, and $m_sm_d/m_b^2$ for Kaon decays. As a result, $K\to \pi\nu\bar\nu$ is practically insensitive to $\delta g_R$, while a mild dependence can be observed in $B_s\to\mu\mu$.
    Overall the constraints are well compatible with each other, with a slight preference for a non-zero $\delta g_R$ driven by the $Z$ decays.
    \item[\textbf{PC}] The case of Partial Compositeness is qualitatively very different. While the left-handed couplings scale in a similar way as in MFV, the light-quark-mass suppression in the right-handed ones is partially lifted by the inverse proportionality to CKM elements. As a consequence, Kaon decays become particularly sensitive to $\delta g_R$.
    This sensitivity is such that a double solution becomes visible, of which one is compatible with the SM, and the other corresponds to an effect which is double in size, but opposite in sign to the SM contribution, effectively giving the same decay rate.
    Both solutions are compatible with $B_s\to\mu\bar\mu$, and within $2\sigma$ with $Z$ decays.
    $B\to K\nu\bar\nu$ can be used here to select one of the solutions, as it pulls towards the SM point.
    Another interesting feature observable in the PC case is the different dependence of the di-neutrino modes on $\delta g_{L(R)}$, with respect to the $B_s$ decay. They are sensitive to orthogonal directions, making  the combination of bounds from all these observables particularly relevant (see Appendix \ref{app:obs} for explicit expressions of the observables).
\end{itemize}

\begin{figure}
    \centering
    \includegraphics[height=0.49\linewidth]{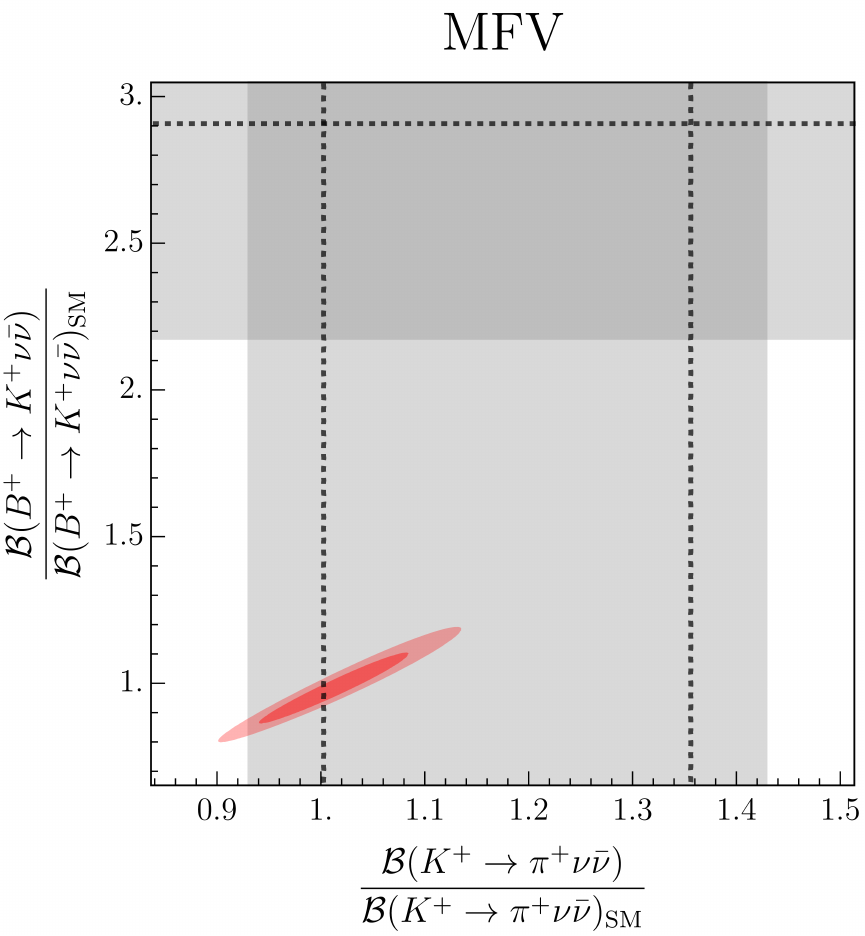}\hspace{2mm
}
    \includegraphics[height=0.49\linewidth]{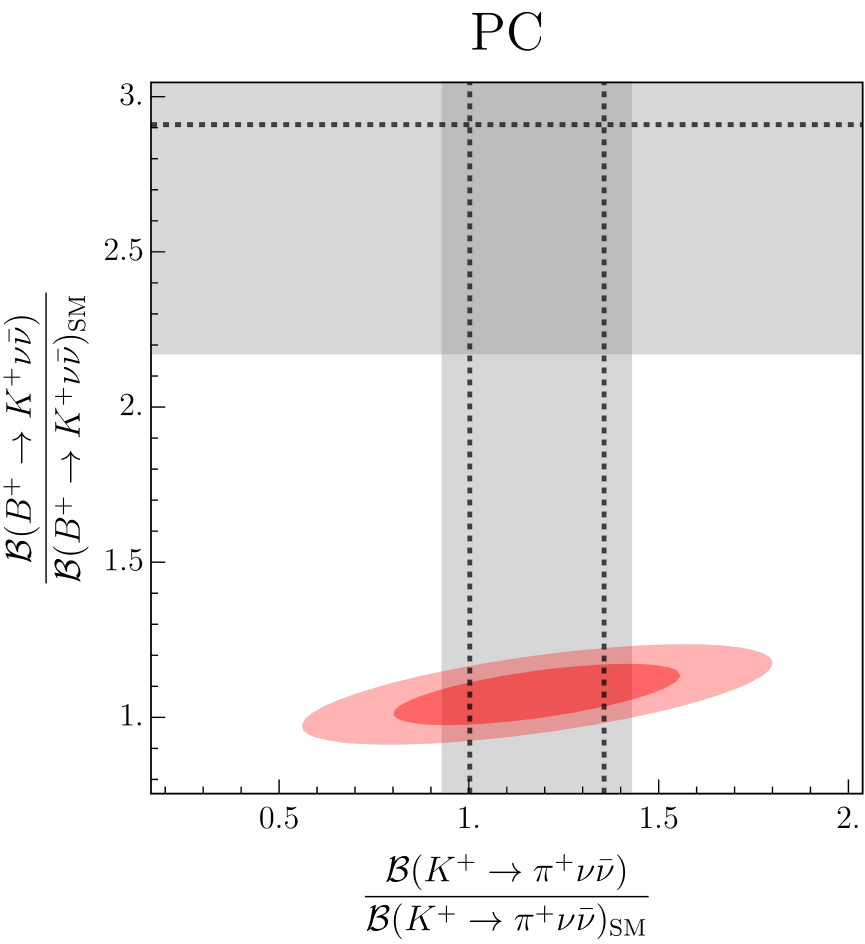}
    \caption{Correlation between $B\to K\nu\bar\nu$ and $K\to\pi\nu\bar\nu$ branching ratios under MFV (left) and PC (right) hypotheses for the $Z$ couplings to down-type quarks. The regions are obtained by varying $\delta g_L$ and $\delta g_R$ around their best-fit values.}
    \label{fig:BKnunupredictions}
\end{figure}

Given our assumptions, only a small set of observables is sufficient to tightly constrain the parameter space.
This implies that other FCNC processes involving down-type quarks are completely fixed, making the framework quite predictive.
In particular, we show in Figure \ref{fig:BKnunupredictions} the relation between $K^+\to\pi^+\nu\bar\nu$ and $B^+\to K^+ \nu\bar\nu$, for both MFV 
and PC, within the allowed regions from the respective fits.
In both cases, the tension with the current experimental value for the $B^+\to K^+\nu\bar\nu$ rate showcases the importance of a more precise measurement for this decay mode. In the plots, the 68\% C.L. ranges of the experimental measurements are shown as grey bands, while dotted lines indicate future NA62 and Belle-II projections, for which we took an expected $15\%$ uncertainty in $K^+\to\pi^+\nu\bar\nu$ and $8\%$ for $B^+\to K^+\nu\bar\nu$ \cite{ATLAS:2025lrr}. In MFV (left plot), as already shown e.g. in~\cite{Bobeth:2005ck,Buras:2013ooa} although with less precise data, an enhancement in $K^+\to\pi^+\nu\bar \nu$ is directly correlated with an enhancement in $B^+\to K^+\nu\bar\nu$, albeit the current experimental precision is still far from the one coming from the global fit (the fit is driven by $B_s\to\mu^+\mu^-$). In PC (right plot), the impact of the Kaon branching ratio is much more evident, as the best-fit region centers around the current experimental value. For this case, smaller effects are expected in $B\to K$ transitions.

\begin{figure}
    \centering
    \includegraphics[width=0.49\linewidth]{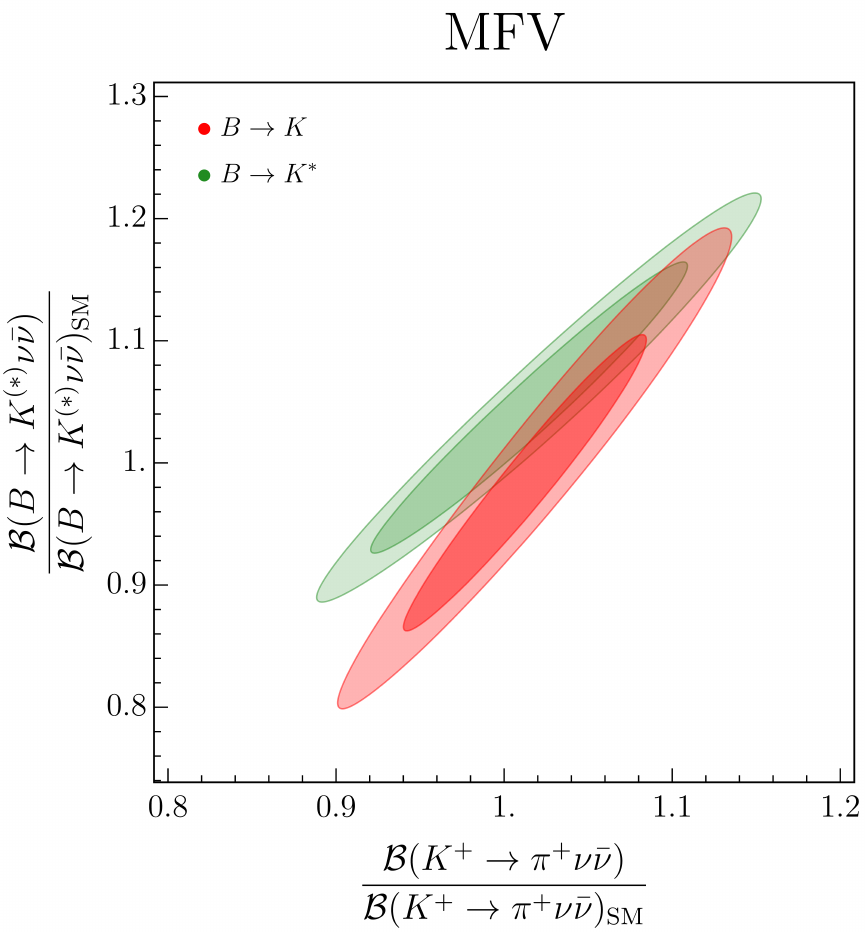}
    \caption{Correlation between $b\to s \nu\bar\nu$ modes and $K^+\to\pi^+\nu\bar\nu$ branching ratio, normalised to their SM expectations. The regions are obtained by varying $\delta g_L$ and $\delta g_R$ around their best-fit values.}
    \label{fig:BKKstvsKpi}
\end{figure}

Finally, we show the difference between the $B^+\to K^+\nu\bar\nu$ and $B\to K^*\nu\bar\nu$ in correlation to $K^+\to \pi^+\nu\bar\nu$ modes in Figure \ref{fig:BKKstvsKpi}, under the MFV hypothesis. The misalignment between the two $b\to s\nu\bar\nu$ modes stems from their different dependence on the right-handed coupling $\delta g_R$, further highlighting the complementarity in measuring both decay rates. The analogous result for the PC case is not shown, as the right-handed coupling is tightly constrained in this case, implying nearly identical correlations for the two $B$ decay modes.

\section{Semileptonic fit with third-generation dominance}
\label{sec:U2}
\subsection{Framework and fit setup}

We now turn to the second scenario, closely following the analysis in Ref.~\cite{Allwicher:2024ncl}.
We work in an EFT framework, using dimension-six semileptonic operators in the SMEFT \cite{Grzadkowski:2010es}, assuming a $U(2)^5$ flavour symmetry acting on the light-fermion families \cite{Barbieri:2011ci}, with NP couplings to third-generation fields only.
Since we are interested in flavour-changing quark transitions, we need to consider breaking terms of the $U(2)_q$ subgroup that acts on the left-handed quark fields only. In this framework, this is effectively achieved by introducing the spurion $\tilde V$, that we parametrise as:
\begin{equation}
    \tilde{V} = -\epsilon \begin{pmatrix}  V_{td} \\
    V_{ts}
    \end{pmatrix}\,.
\end{equation}
with $\epsilon$ being a positive, $\mathcal{O}(1)$ parameter. Furthermore, we have to specify the alignment in flavour space of the the quark left-handed doublets. We choose a down-aligned basis, such that $q^i_L = (V^*_{ji}u^j_L, d^i_L)^T$, with $V$ being the CKM matrix. With this, the $U(2)_q$-invariant building blocks are the third-generation doublet $q^3_L$, and the combination $\tilde{V}_i q^i_L$. In the lepton sector, we consider only third-generation fields as $\ell^3_L = (\nu_\tau,\tau_L)$ and $\tau_R$. \\
With these elements, the effective Lagrangian we consider is
\begin{equation}
\label{eq:lagrangian}
\mathcal{L}^{\rm NP}_{\mathrm{eff}}\supset \sum_k C_k Q_k+{\rm h.c.}\,,
\end{equation}
where we consider the operators
\begin{equation}
Q_{\ell q}^{\pm}=(\bar{q}_L^3 \gamma^{\mu} q_L^3)(\bar{\ell}_L^3\gamma_{\mu}\ell_L^3)  \pm ( \bar{q}_L^3 \gamma^{\mu}\sigma^a q_L^3)(\bar{\ell}_L^3\gamma_{\mu}\sigma^a\ell_L^3)\,,\quad
Q_{S}=(\bar{\ell}_L^{3}\tau_R)(\bar{b}_R q_L^3)\,.
\label{eq:ops}
\end{equation}
and the Wilson coefficients $C_{\ell q}^{}\pm$ and $C_S$ are always taken real, and expressed in units of $\mathrm{TeV}^{-2}$.
 A priori, in addition to the operators in Eq.~\eqref{eq:ops}, one could construct all possible combinations obtained by replacing $q^3_L\to \tilde{V}_{i} q_L^i$ once or twice in the quark bilinear. Each replacement is effectively a different effective operator and, therefore, weighted by an independent Wilson coefficient. However, in order to obtain a predictive framework, we adopt a minimal ansatz,  replacing
\begin{equation}
    q^3_L\to q^3_L+\tilde{V}_{i} q_L^i\,.
\end{equation}
With this choice, the relative size of operators involving light quark generations with respect to those involving only the third generation is entirely controlled by the spurion $\tilde{V}$. 

Under these assumptions, our framework is completely determined by four independent parameters: $C^{+}_{\ell q}$,$C^{-}_{\ell q}$, $C_S$, and $\epsilon$, which have to be determined from data. In our analysis, we consider the same observables as in \cite{Allwicher:2024ncl}. They include LHC Run-II di-tau and mono-tau data \cite{Allwicher:2022mcg}, electroweak observables, lepton-flavour universality ratios in $\tau$ decays\cite{HFLAV:2024ctg}, $R_{D^{(*)}}$\cite{HFLAV:2024ctg}, $B\to K\nu\bar\nu$\cite{Belle-II:2023esi,Belle:2017oht}, $K^+\to\pi^+\nu\bar\nu$ and RG-induced contributions to $b\to\ s\mu^+\mu^-$ \cite{NA62:2026rwr} from $(\bar{s}_L\gamma^\mu b_L)(\tau_L\gamma_\mu \tau_L)$ \cite{Aebischer:2017gaw,Crivellin:2018yvo,Aebischer:2022oqe}. With respect to Ref.~\cite{Allwicher:2024ncl}, we update two experimental inputs. First, we use the most recent result for the branching fraction $\mathcal{B}(K^+\to\pi^+\nu\bar\nu)$ from the NA62 Collaboration \cite{NA62:2026rwr}. Second, we update the short-distance input entering $b\to s\mu^+\mu^-$ using the analysis in Ref.~\cite{Bordone:2026wov}. For all other inputs, as well as for the EFT expansion of the observables included in the fit, we refer to Ref.~\cite{Allwicher:2024ncl}. 

\subsection{Fit results and implications}
\begin{figure}
\centering\includegraphics[width=0.5\linewidth]{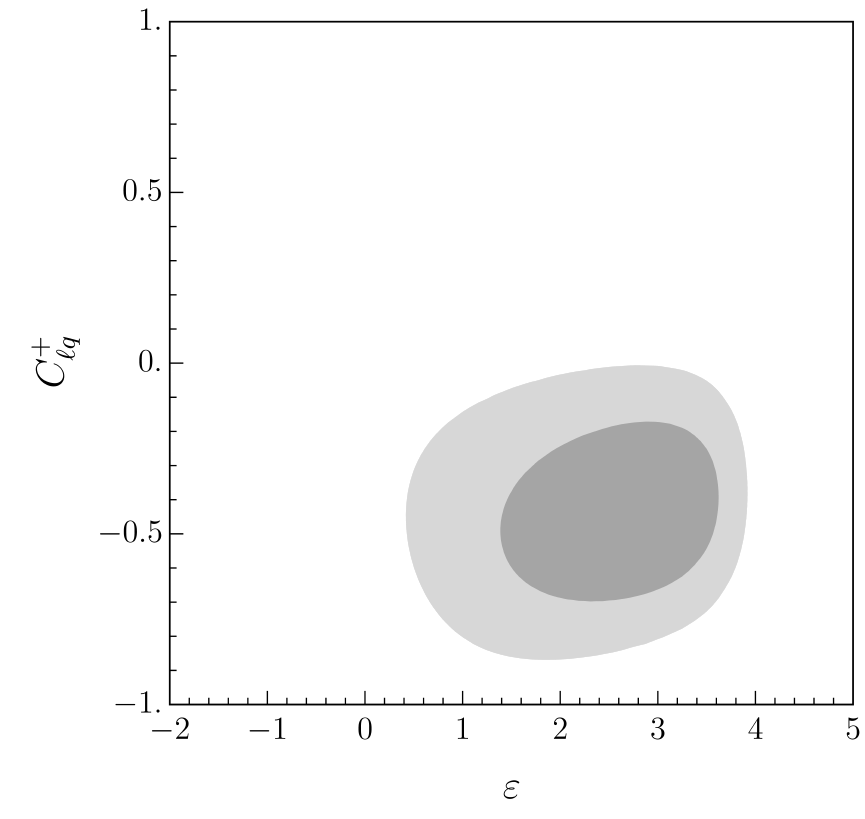}\includegraphics[width=0.5\linewidth]{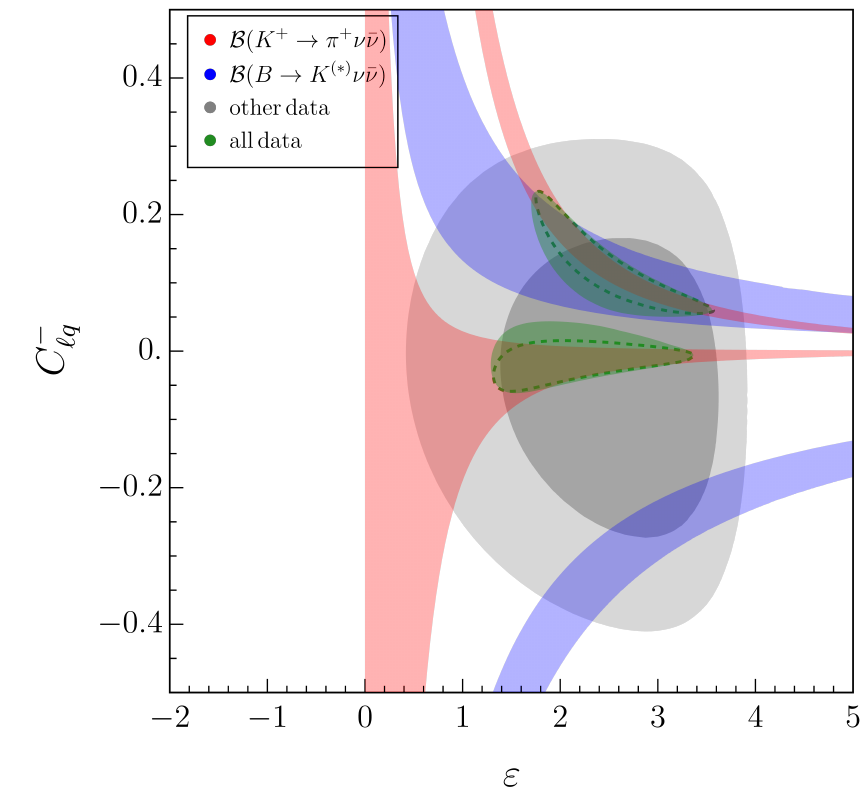}
    \caption{Profiled constraints in the $C_{\ell q}^+$--$\epsilon$ plane (left) and in the $C_{\ell q}^-$--$\epsilon$ plane (right).
The dark- and light-grey regions denote the $68\%$ and $95\%$ preferred regions from the global fit without di-neutrino observables.
The blue and red bands show the $68\%$ regions allowed by $B\to K^{(*)}\nu\bar\nu$ and $K^+\to\pi^+\nu\bar\nu$, respectively.
The green regions indicate the $68\%$ preferred regions of the combined fit including all observables, while the dashed green contours show the corresponding projection assuming the expected final NA62 sensitivity, with a relative uncertainty of $15\%$ on $\mathcal{B}(K^+\to\pi^+\nu\bar\nu)$.
}
    \label{fig:cminusplus}
\end{figure}
We perform a fit to the aforementioned data to obtain the best fit region for the fit parameters. We first comment on $C_S$. This Wilson coefficient is very well  constrained by high-$p_T$ searches and is found to be compatible with zero. Therefore, in the following, we focus on $C_{\ell q}^+$, $C_{\ell q}^-$ and $\epsilon$ only.

The results are shown in Figure~\ref{fig:cminusplus}. In both panels, the dark- and light-grey regions denote the $68\%$ and $95\%$ preferred regions obtained from the global fit without di-neutrino observables, while the red and blue bands show the $68\%$ regions allowed by $K^+\to\pi^+\nu\bar\nu$ and $B\to K^{(*)}\nu\bar\nu$, respectively. Note that since the modification to the $B\to K^{(*)}\nu\bar\nu$ is SM-like, no recasting as presented in \cite{Gartner:2024muk} is necessary. In each panel, all remaining fit parameters are profiled over.  In the left panel we show the two-dimensional distribution in the $C^+_{\ell q}$--$\epsilon$ plane. Neglecting small RGE effects, this plane is unaffected by the di-neutrino modes. With respect to our previous analysis \cite{Allwicher:2024ncl}, we observe some changes due to the reduced deviation in $b\to s\mu^+\mu^-$ data and the increased error on them, resulting in a more conservative description of non-local effects\cite{Bordone:2026wov}. As a consequence, the preferred central value of $C_{\ell q}^+$ is shifted towards smaller values, while its relative uncertainty increases.\\
In the right panel of Figure~\ref{fig:cminusplus}, we show the corresponding $C_{\ell q}^-$--$\epsilon$ plane. As in the case of $C_{\ell q}^+$, the preferred region without di-neutrino modes is slightly shifted towards lower values of $C_{\ell q}^-$ compared to \cite{Allwicher:2024ncl}, even though the shift is less pronounced. The impact of the di-neutrino modes is shown by the blue and red bands, corresponding to the $68\%$ regions allowed by $B\to K^{(*)}
\nu\bar\nu$ and $K^+\to\pi^+\nu\bar\nu$, respectively. The new measurement of $\mathcal{B}(K^+\to\pi^+\nu\bar\nu)$, which has a significantly smaller uncertainty than the previous one, considerably reduces the width of the allowed region overlapping with the $B^+\to K^+\nu\bar\nu$ mode. Given the agreement of this measurement with the SM prediction, the solution with $\epsilon\sim 0$ or $C_{\ell q}^-\sim 0$ now lies within the $68\%$ allowed region. This region is, however, in mild tension with other low-energy observables, specifically $R_{D^{(*)}}$ and $B\to K^{(*)}\nu\bar\nu$, which show some deviations with respect to the corresponding SM predictions. For this reason, the most interesting branch lies in the upper-right part of the right panel of Figure~\ref{fig:cminusplus}. This red branch corresponds to a large BSM contribution, with opposite sign with respect to the SM one. The two contributions therefore compensate each other, effectively generating a small net shift in $\mathcal{B}(K^+\to\pi^+\nu\bar\nu)$.  The preferred region obtained from the fit to all data is shown by the green $68\%$ green areas. Despite the fact that the two green lobes are equally allowed due to the large experimental uncertainties, in the following we focus on the upper lobe, which maximises the possible BSM effects in the observables discussed below. Finally, the dashed green contours show how the preferred region would change under the projected NA62 sensitivity, assuming the current central value and an expected relative uncertainty of $15\%$ on the branching fraction. The reduction in the size of the allowed region, compared to the current one, highlights the significant potential of more precise measurements of this decay.

\section{Discussion}\label{sec:KL}

The two scenarios presented in the previous Sections exhibit distinct phenomenological features. First of all, low-energy observables allow us to extract the effective NP scales associated with these scenarios by matching their contributions onto the relevant four-fermion operators. 
In the case of the modified $Z$ couplings, we find
\begin{equation}
    \Lambda_\mathrm{eff}^{Z} \gtrsim 8\,\mathrm{TeV}\,,
\end{equation}
for both MFV and PC. This similarity arises because, at low energies, the dominant constraint is on $\delta g_L$, whose size is comparable for both flavour assumptions, while $\delta g_R$ appears in low-energy flavour observables through a suppressed contribution. We also note that measurements of FCNC-mediated processes, with the exception of $B^+\to K^+\nu\bar\nu$, are in good agreement with the SM. This leaves comparatively little room for NP to give large contributions, which therefore translates into a higher effective NP scale. Even though this effective scale is rather high, the tree-level couplings to the valence quarks make it accessible to the HL-LHC programme.

The case of semileptonic effective operators with third-generation dominance is qualitatively very different. In this case in fact, the addition of more semileptonic observables, especially the lepton-flavour universality ratio $R_{D^{(*)}}$, makes NP contributions larger. In addition, the constraint from $B_s\to\mu^+\mu^-$, crucial in the modified $Z$-coupling case, falls short because of the third-generation hypothesis. In this case, we find that the effective scale is
\begin{equation}
    \Lambda_\mathrm{eff}^{U(2)} \gtrsim 1.8\,\mathrm{TeV}\,,
\end{equation}
considerably lower than in the modified-$Z$ scenarios and, therefore, much closer to the energy scales directly explored at the LHC. It is also worth stressing that this picture would not be qualitatively altered by the inclusion of additional operators that, for example, generate tree-level contributions to $B_s\to\mu^+\mu^-$, as in the modified-$Z$-couplings scenario. In the $U(2)^5$ picture, in fact, such contributions would introduce either an additional Wilson coefficient or a new spurion in the lepton sector. Since these new parameters are largely independent of those entering the fit of Sec.~\ref{sec:U2}, they would not significantly modify the constraints presented here \cite{Covone:2025lee}.

Finally, we further comment on the $K^+\to\pi^+\nu\bar\nu$ mode, along with $K_L\to\pi^0\nu\bar\nu$. For what concerns the modified $Z$ couplings scenario, $K^+\to\pi^+\nu\bar\nu$ and $K_L\to\pi^0\nu\bar\nu$ are predicted to have similar size. A large misalignment between the two would require, for example, a large complex component in $\delta g_L^{21}$ and/or $\delta g_R^{21}$, as illustrated in \cite{Blanke:2026rdi}. Currently, constraints on the imaginary parts of $\delta g_{L,R}^{21}$ are weak. Therefore, without a measurement of $K_L\to\pi^0\nu\bar\nu$, we refrain from making any assumption about their size, and assume these couplings to be real. With this, we find that both decays are predicted to be largely SM-like under both the MFV and PC hypotheses.

\begin{figure}
\centering
\includegraphics[width=0.49\linewidth]{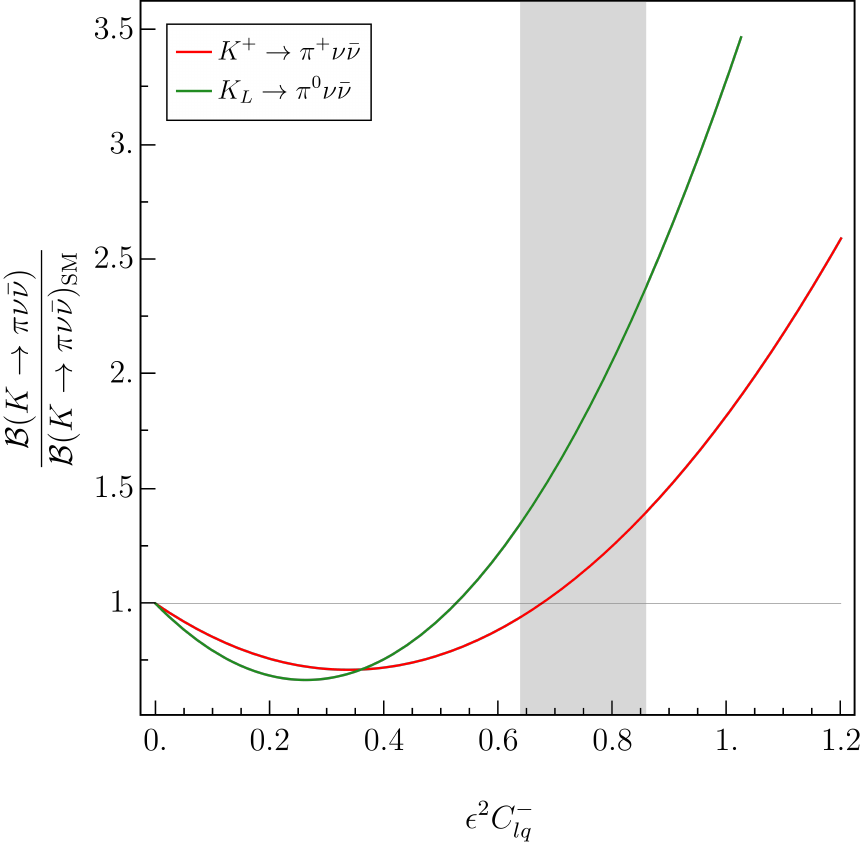}
\includegraphics[width=0.49\linewidth]{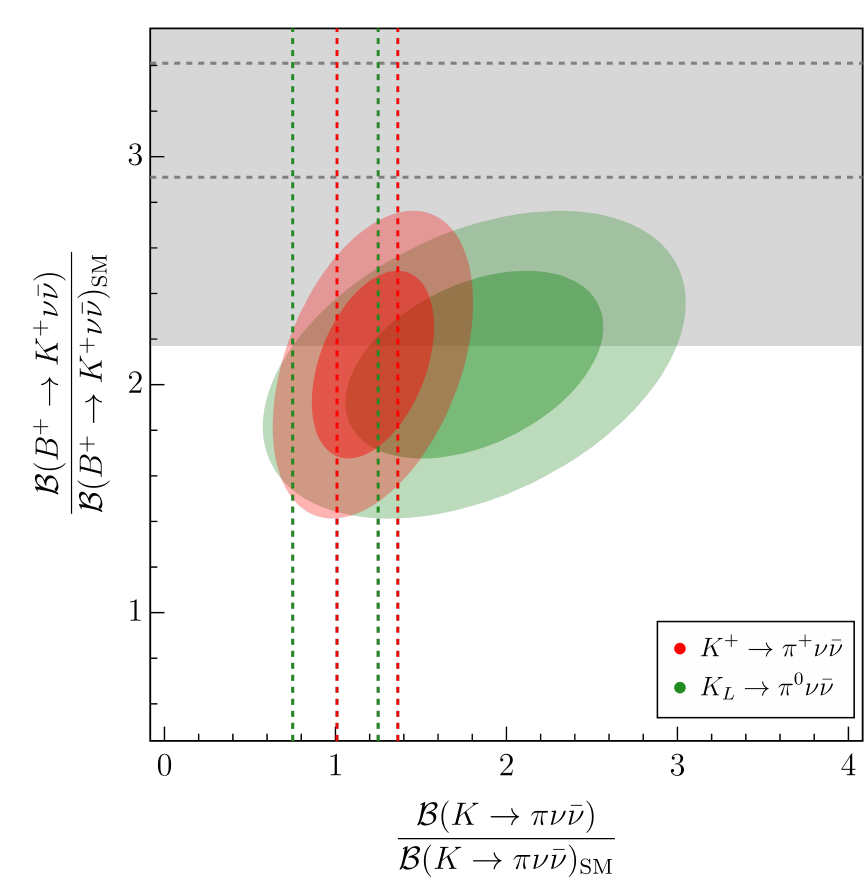}
    \caption{Effects in the two $K\to\pi\nu\bar\nu$ modes in the $U(2)$ semileptonic scenario. On the left, predictions of the decay rates as a function of the Wilson coefficient $C_{lq}^-$, with the current best-fit interval (68\% C.L.) shown as a grey band. On the right, different correlations of the two modes with the $B^+\to K^+\nu\bar\nu$ decay rate, in the current global fit. The coloured dashed lines indicate future projections from NA62 and KOTO-II, while the grey band and dashed lines show the current and projected sensitivities from Belle II.}
    \label{fig:U2_pheno}
\end{figure}

In the semileptonic case, instead, the situation is again quite different. The branching ratios for $K^+\to\pi^+\nu\bar\nu$ and $K_L\to\pi^0\nu\bar\nu$ read

\begin{equation}
\begin{aligned}
     \mathcal{B}(K^+\to\pi^+\nu\bar\nu)
      &= 2 \mathcal{B}(K^+\to\pi^+\nu_e\bar\nu_e)_\mathrm{SM} +\mathcal{B}(K^+\to\pi^+\nu_\tau\bar\nu_\tau)_\mathrm{SM}\left| 1
      + \varepsilon^2 \frac{\pi v^2}{\alpha} 
      \frac{C_{\ell q}^{-}}{ C_{sd,\tau}^\mathrm{SM}}\right|^2\,, \\
      &= 2 \mathcal{B}(K^+\to\pi^+\nu_e\bar\nu_e)_\mathrm{SM} +\mathcal{B}(K^+\to\pi^+\nu_\tau\bar\nu_\tau)_\mathrm{SM}\left( 1
      - 2.91 \varepsilon^2 C_{\ell q}^{-}\right)^2\,, \\
      \mathcal{B}(K_L\to\pi^0\nu\bar\nu)
      &= \mathcal{B}(K_L\to\pi^0\nu\bar\nu)_\mathrm{SM}\left[\frac{2}{3}+\frac{1}{3}\bigg( 1
      + \varepsilon^2 \frac{\pi v^2}{\alpha} 
      \frac{C_{\ell q}^{-}}{C_{sd,K_L}^\mathrm{SM}}\bigg)^2\right]\,, \\
      &= \mathcal{B}(K_L\to\pi^0\nu\bar\nu)_\mathrm{SM}\left[\frac{2}{3}+\frac{1}{3}\bigg( 1
      -3.82 \varepsilon^2 
      C_{\ell q}^{-}\bigg)^2\right]\,,
\end{aligned}
\end{equation}
with $C_{sd,\tau}^\mathrm{SM}$ as in Eq.~(\ref{eq:WCKpi}) and $C_{sd,K_L}^{\textrm{SM}}\sim -X_t/s_W^2$, from which we can read the different dependence on $ \epsilon^ 2C_{lq}^-$ of the two modes. The different numerical coefficients multiplying the $ \epsilon^ 2C_{lq}^-$ term for the $K^+\to\pi^+\nu\bar\nu$ and $K_L\to\pi^0\nu\bar\nu$ modes determine whether the predicted branching fractions lie above or below their SM values, as well as which of the two modes exhibits the larger relative enhancement or suppression. This pattern is illustrated in Figure~\ref{fig:U2_pheno}.
In the left panel, we show exactly the correlation between the $K^+\to\pi^+\nu\bar\nu$ and $K_L\to\pi^0\nu\bar\nu$ modes, as a function of the combination $\epsilon^2 C^-_{\ell q}$. Depending on the value of this combination, the two modes follow a specific pattern, with one branching fraction being modified relative to the SM more strongly
than the other. The preferred $1\sigma$ fit region is highlighted in gray. Within this region, there is a clear preference for the enhancement of $K_L\to\pi^0\nu\bar\nu$ branching fraction to be larger than for the  $K^+\to\pi^+\nu\bar\nu$, 
with 
\begin{equation}
 \frac{
   \mathcal{B}(K_L\to\pi^0\nu\bar\nu)/
   \mathcal{B}(K_L\to\pi^0\nu\bar\nu)_{\rm SM}
 }{
   \mathcal{B}(K^+\to\pi^+\nu\bar\nu)/
   \mathcal{B}(K^+\to\pi^+\nu\bar\nu)_{\rm SM}
 }
 \sim 1.5\,.
\end{equation}
The corresponding predictions remain well within the Grossman--Nir bound
\cite{Grossman:1997sk}. This enhancement is a direct phenomenological consequence of this scenario, which can be tested by future measurements of the two modes with
improved precision. In the right panel of Figure~\eqref{fig:U2_pheno}, we show the correlation between the $B^+\to K^+\nu\bar\nu$ and $K^+\to\pi^+\nu\bar\nu$ branching fraction in red and the $K_L\to\pi^0\nu\bar\nu$ in green, with all branching fractions normalised to the respective SM prediction. The dark and light colored red (green) regions are the $68\%$ and $95\%$ confidence regions, respectively, while the gray area is the experimental region currently allowed by $B^+\to K^+\nu\bar\nu$. We stress that the allowed region for the $\mathcal{B}(K_L\to\pi^0\nu\bar\nu)$ is a prediction of our framework. Here, the  cancellation in the $K^+\to\pi^+\nu\bar\nu$ mode discussed above is clearly visible, with the best fit region containing the SM prediction for this mode, while still allowing for sizeable deviations in $B^+\to K^+\nu\bar\nu$. By contrast, larger effects are allowed in $K_L\to\pi^0\nu\bar\nu$. Nevertheless, at the present level
of precision, the predictions for the two Kaon modes remain compatible
within the $1\sigma$ region.

The projected final NA62 sensitivity of $15\%$ and the projected KOTO-II $25\%$ sensitivity  \cite{KOTO:2024zbl} are shown with dashed red and green lines, respectively. Since no measurement of the $\mathcal{B}(K_L\to\pi^0\nu\bar\nu)$ is currently available, the projected KOTO-II region is centered around the SM. Together with the expected $8\%$ precision
on the $B^+\to K^+\nu\bar\nu$ branching fraction \cite{ATLAS:2025lrr}, shown by the dashed gray lines, measurements of all three modes would provide a powerful test of this framework. Should the present central values remain unchanged, the improved experimental precision would challenge this scenario. Such a tension  could be leveraged by non-minimal breaking of the $U(2)^5$ flavour symmetry, introducing a further misalignment between the two components of the spurion $\tilde{V}$, as illustrated in Ref.~\cite{Allwicher:2024ncl}.

\section{Conclusions}
The recent NA62 measurement of $K^+\to\pi^+\nu\bar\nu$, with a relative uncertainty below $20\%$, brings FCNC-mediated Kaon decays into the precision era. While this represents a remarkable experimental achievement and the result is compatible with the corresponding SM prediction, the present precision still leaves room for sizeable new-physics effects, depending on the underlying flavour hypothesis. Our analysis addresses precisely this point by studying the resulting correlations among Kaon and $B$-meson observables.

We investigated two complementary scenarios.
We first focussed on modified $Z$-coupling scenarios, which, by construction, affect FCNC-mediated processes only. In this case, we introduced two distinct flavour hypotheses, Minimal Flavour Violation and Partial Compositeness. In the former, the $K^+\to\pi^+\nu\bar\nu$ constraints are weaker than the ones from $B_s\to\mu^+\mu^-$, but this hierarchy is  reversed in the PC case,  where $K^+\to\pi^+\nu\bar\nu$ provides an orthogonal constraint to the $B_s\to\mu^+\mu^-$ mode. In both cases, we predict the $B^+\to K^+\nu\bar\nu$ branching ratio, whose current experimental central value lies well above our predictions. This highlights the complementarity of these observables and the importance of measuring them precisely to test possible new-physics scenarios. We also predicted the $B\to K^*\nu\bar\nu$ branching fraction, showing that combined measurements of $B\to K^*\nu\bar\nu$ and $B^+\to K^+\nu\bar\nu$ can help to further constrain the allowed NP parameter space.

In contrast, the semileptonic scenario exhibits a different pattern of correlations, arising from the interplay between a larger set of low energy flavour-changing observables, electroweak precision tests, and high-$p_T$ data. After imposing all available constraints, the branching fractions of $B\to K^{(*)}\nu\bar\nu$, $K^+\to\pi^+\nu\bar\nu$ and $K_L\to\pi^0\nu\bar\nu$ become correlated predictions, with the latter being a genuine prediction of the framework. Future measurements of these modes will therefore offer a powerful test of our hypothesis.

\section*{Acknowledgments}
We thank Gino Isidori for suggesting this projects and for useful discussions. The work of M.B. is supported by the Cluster of Excellence \textit{PRISMA}$^{++}$ (EXC 2118/2) funded by the German Research Foundation (DFG) under Germany’s Excellence Strategy (Project ID 390831469). LA acknowledges funding from the Deutsche
Forschungsgemeinschaft under Germany’s Excellence Strategy EXC 2121 “Quantum Universe” – 390833306, as well as from the grant 491245950.

\appendix

\section{Low-energy observables}\label{app:obs}

In this section we provide details on the low-energy observables as functions of the modified $Z$ couplings. For expressions in terms of the semileptonic LEFT and SMEFT Wilson coefficients see e.g. \cite{Allwicher:2024ncl}.

\subsection{$\bar{B}_s\to\mu^+\mu^-$}
At low energies, the effective Lagrangian describing $b\to s\ell^+\ell^-$ transitions reads
\begin{equation}
    \mathcal{L}(b\to s\ell^+\ell^-) = \frac{4 G_F}{\sqrt{2}} V_{tb}V_{ts}^*\frac{\alpha}{4\pi} \left[C_9 \mathcal{O}_9+C^\prime_9 \mathcal{O}^\prime_9+C_{10} \mathcal{O}_{10}+C^\prime_{10}\mathcal{O}^\prime_{10}\right]\,,
\end{equation}
where
\begin{equation}
    \mathcal{O}_9^{(\prime)} = (\bar{s}\gamma_\mu P_{L(R)} b)(\bar{\ell}\gamma^\mu \ell)\,, \qquad \mathcal{O}_{10}^{(\prime)} = (\bar{s}\gamma_\mu P_{L(R)} b)(\bar{\ell}\gamma^\mu\gamma_5 \ell)\,,
\end{equation}
and $C_{9(10)}^{\prime}=0$ in the SM and we parametrise $C_{9(10)} = C_{9(10)}^\mathrm{SM}+ \Delta C_{9(10)}$.
We have that
\begin{align}
    \Delta C_9 =&\,+\frac{\sqrt{2}\pi^2}{G_F V_{tb} V_{ts}^* m_W^2 s_W^2}\delta g_L^{23}\,,\\
    \Delta C_{10} =&\, -\frac{\sqrt{2}\pi^2}{G_F V_{tb} V_{ts}^* m_W^2 s_W^2}\delta g_L^{23}\,,\\
    C^\prime_9 =&\, +\frac{\sqrt{2}\pi^2}{G_F V_{tb} V_{ts}^* m_W^2 s_W^2}\delta g_R^{23}\,,\\
    C^\prime_{10} =&\,-\frac{\sqrt{2}\pi^2}{G_F V_{tb} V_{ts}^* m_W^2 s_W^2}\delta g_R^{23}\,, 
\end{align}
The branching fraction of $\bar{B}_s\to\mu^+\mu^-$ depends on $C_{10}$ and $C_{10}^\prime$:
\begin{equation}
\begin{aligned}
    \mathcal{B}(\bar{B}_s\to\mu^+\mu^-) =\,& \mathcal{B}(\bar{B}_s\to\mu^+\mu^-)_\mathrm{SM}\bigg| 1+ \frac{\Delta C_{10}-C_{10}^\prime}{C_{10}^\mathrm{SM}}\bigg|^2 \\
    =\,& \mathcal{B}(\bar{B}_s\to\mu^+\mu^-)_\mathrm{SM}\bigg|1+ \frac{\sqrt{2}\pi^2}{G_F V_{tb} V_{ts}^* m_W^2 Y(x_t)}(\delta g_L^{23}-\delta g_R^{23})\bigg|^2\,,
\end{aligned}
\end{equation}
where $Y(x_t)\approx 0.957$.
For the numerical analysis, we use the combination of the latest measurements from LHCb and CMS \cite{LHCb:2021awg,LHCb:2021vsc,CMS:2022mgd} in \cite{Greljo:2022jac}, and the QED-improved SM prediction in \cite{Beneke:2017vpq,Beneke:2019slt}. The values read:
\begin{equation}
    \mathcal{B}(\bar{B}_s\to\mu^+\mu^-)_\mathrm{exp}= 3.32^{+0.32}_{-0.25}\times 10^{-9}\,, \qquad \mathcal{B}(\bar{B}_s\to\mu^+\mu^-)_\mathrm{SM}=(3.66 \pm 0.14)\times 10^{-9}\,.
\end{equation}

\subsection{$K\to\pi\nu\bar\nu$}
The low-energy effective Lagrangian contributing to $s\to d\nu\bar\nu$ transition is
\begin{equation}
    \mathcal{L}(s\to d\nu\bar\nu) = -\frac{4G_F}{\sqrt{2}} \frac{\alpha}{2\pi} V_{ts}^* V_{td} \left[C^{sd,\ell}_L \mathcal{O}_L+C^{sd,\ell}_R \mathcal{O}_R\right]\,,
\end{equation}
where
\begin{equation}
    \mathcal{O}_{L(R)} = (\bar{s}\gamma_\mu P_{L(R)}d)(\bar{\nu}_\ell\gamma^\mu P_L \nu_\ell)\,.
\end{equation}
The right-handed Wilson coefficient is zero in the SM, while, for the left-handed part, we use
\begin{equation}
C^{sd,\ell}_L = C_{sd,\ell}^\mathrm{SM}+\Delta C_L^{sd}\,, \quad \textrm{and} \quad C_{sd,\ell}^\mathrm{SM} = -\frac{1}{s_W^2} \left(X_t + \frac{V_{cs}^* V_{cd}}{V_{ts}^* V_{td}} X_c^\ell\right)\,,
\label{eq:WCKpi}
\end{equation}
where we employ $X_t=1.462$, $X_c^e = X_c^\mu = 1.053\cdot 10^{-3}$ and $X_c^\tau = 0.711\cdot 10^{-3}$. We find
\begin{align}
    \Delta C_L^{sd,\ell} =\,& -C_{sd,\ell}^\mathrm{SM}\frac{\sqrt{2}\pi^2}{G_F m_W^2 V_{ts}^* V_{td}\left(X_t + \frac{V_{cs}^* V_{cd}}{V_{ts}^* V_{td}} X_c^\ell\right)}\delta g_L^{21}\,, \\
    C_R^{sd,\ell} =\,& -C_{sd,\ell}^\mathrm{SM}\frac{\sqrt{2}\pi^2}{G_F m_W^2V_{ts}^* V_{td}\left(X_t + \frac{V_{cs}^* V_{cd}}{V_{ts}^* V_{td}} X_c^\ell\right)}\delta g_R^{21}\,.
\end{align}
This gives
\begin{equation}
\begin{aligned}
    \mathcal{B}(K^+\to\pi^+\nu\bar\nu) =\,& \mathcal{B}(K^+\to\pi^+\nu_e\bar\nu_e)_\mathrm{SM}\sum_{\ell = e,\mu}\left|1-\frac{\sqrt{2}\pi^2 (\delta g_L^{21}+\delta g_R^{21})}{G_F m_W^2 V_{ts}^* V_{td}\left(X_t + \frac{V_{cs}^* V_{cd}}{V_{ts}^* V_{td}} X_c^\ell\right)}\right|^2\\
    +&\mathcal{B}(K^+\to\pi^+\nu_\tau\bar\nu_\tau)_\mathrm{SM}\left|1-\frac{\sqrt{2}\pi^2 (\delta g_L^{21}+\delta g_R^{21})}{G_F m_W^2 V_{ts}^* V_{td}\left(X_t + \frac{V_{cs}^* V_{cd}}{V_{ts}^* V_{td}} X_c^\tau\right)}\right|^2\,,
    \end{aligned}
\end{equation}
with $\mathcal{B}(K^+\to\pi^+\nu_e\bar\nu_e)_\mathrm{SM} = 2.87 \cdot 10^{-11} $ and $\mathcal{B}(K^+\to\pi^+\nu_\tau\bar\nu_\tau)_\mathrm{SM} = 2.35 \cdot 10^{-11}$. The corresponding SM prediction \cite{Allwicher:2024ncl}, based on \cite{Buchalla:1998ba,Brod:2008ss,Buras:2006gb,Brod:2010hi,Brod:2021hsj,Isidori:2005xm,Lunghi:2024sjy,Mescia:2007kn}, and experimental measurement \cite{NA62:2026rwr} that we use in our analysis are:
\begin{equation}
    \mathcal{B}(K^+\to\pi^+\nu\bar\nu)_\mathrm{SM} = (8.09 \pm 0.63)\times 10^{-11}\qquad \mathcal{B}(K^+\to\pi^+\nu\bar\nu)_\mathrm{exp} = 9.6^{+1.9}_{-1.8}\times 10^{-11}
\end{equation}
Similarly, one obtains for the $K_L \to \pi^0 \nu\bar\nu$ decay mode
\begin{align}
    \mathcal{B}(K_L\to\pi^0\nu\bar\nu) =\,& \mathcal{B}(K_L\to\pi^0\nu\bar\nu)_\mathrm{SM} \left(1-\frac{\sqrt{2}\pi^2}{G_F m_W^2 X_t} \frac{{\rm Im} (\delta g_L^{21}+\delta g_R^{21})}{{\rm Im}(V_{ts}^* V_{td})}\right)^2 \,,
\end{align}
where the SM prediction is
\begin{equation}
    \mathcal{B}(K_L\to\pi^0\nu\bar\nu)_\mathrm{SM} = (2.58 \pm 0.30) \times 10^{-11} \,.
\end{equation}

\subsection{$B\to K^{(*)} \nu\bar\nu$}

The effective Lagrangian for $B\to K \nu\bar\nu$ at low energies is the same as the one for $K\to\pi\nu\bar\nu$ transitions, with the replacement $d\to b$ in the effective operator and the CKM matrix elements.
The contribution from charm quarks in the SM can be safely neglected in this case, and the expression including new physics effects simplifies to
\begin{align}
    \mathcal{B} (B^+\to K^+\nu\bar\nu) = \mathcal{B} (B^+\to K^+\nu\bar\nu)_{\rm SM} \left|1 - \frac{\sqrt{2}\pi^2 (\delta g_L^{23}+\delta g_R^{23})}{G_F m_W^2 V_{ts}^* V_{tb}X_t}\right|^2 \,.
\end{align}
For the $B\to K^*$ mode, an additional contribution from the axial current arises, yielding
\begin{align}
    \mathcal{B} (B\to K^*\nu\bar\nu) &= \mathcal{B} (B\to K^*\nu\bar\nu)_{\rm SM}\Bigg{\{} \left|1 - \frac{\sqrt{2}\pi^2 (\delta g_L^{23}+\delta g_R^{23})}{G_F m_W^2 V_{ts}^* V_{tb}X_t}\right|^2 \\
    &+ \, \eta_{K^*} {\rm Re}\left[\frac{\sqrt{2}\pi^2}{G_F m_W^2 V_{ts}^* V_{tb} X_t} \delta g_R^{23}\left(1-\frac{\sqrt{2}\pi^2}{G_F m_W^2 V_{ts}^* V_{tb} X_t} \delta g_L^{23}\right)\right] \Bigg{\}} \,,
\end{align}
with $\eta_{K^*} = 3.33$.
The SM predictions read \cite{Becirevic:2023aov}
\begin{align}
    \nonumber
    \mathcal{B} (B^+\to K^+\nu\bar\nu)_{\rm SM} &= (4.72 \pm 0.27)\times 10^{-6} \,, \\
    \mathcal{B} (B^0\to K^{*0}\nu\bar\nu)_{\rm SM} &= (9.7 \pm 1.4)\times 10^{-6} \,,
\end{align}
based on \cite{Buras:2014fpa,Altmannshofer:2009ma,Buchalla:1992zm,Buchalla:1998ba,Misiak:1999yg,Brod:2010hi}.

\bibliographystyle{utphys}
\bibliography{refs}

\end{document}